\documentstyle[aps,multicol]{revtex}

\begin{document}

\draft
%\tightenlines
 
\title{On the effective interactions 
between like-charged macromolecules}

\author{Emmanuel Trizac
\footnote{Electronic Address: Emmanuel.Trizac@th.u-psud.fr}}

\address{
Laboratoire de Physique Th\'eorique (Unit\'e Mixte de Recherche UMR 8627 du CNRS), \\
B\^atiment 210, Universit\'e de Paris-Sud, 91405 Orsay Cedex, France
}
 
\date{\today}

\maketitle
\begin{abstract}
We investigate, within a local density functional theory formalism, 
the interactions between like-charged polyions immersed in a
confined electrolyte. We obtain a simple condition for a 
repulsive effective pair potential,
that can be related to the thermodynamic stability criterion of the
uncharged  counterpart of microscopic species constituting the electrolyte. 
Under the same condition, the phenomenon of charge inversion
(over-charging), where the polyion bare charge is over-screened by its
electric double layer, is shown to be impossible. These results hold
beyond standard mean-field theories (such as Poisson-Boltzmann or Modified
Poisson-Boltzmann approaches).

\end{abstract}


%\begin{multicols}{2}

Electrostatic forces play a key role in determining the stability, 
phase and structural properties of colloids or macroions suspended in polar
solvents (usually aqueous media). Such suspensions are ubiquitous
in many biological and technologically important systems, from
superabsorbants or water-soluble paints to DNA solutions. 
Their behavior is however far from being well understood, despite an intense 
theoretical effort over the last fifty years. 

Upon dispersion in water, mesoscopic colloidal polyions bearing ionizable groups
gain a (large) bare charge, and strongly attract counterions 
while repelling coions, thus giving rise to electric double layers characterized
by strong inhomogeneities in the local density of these
microions in the vicinity of the polyions. 
From the electrostatic part of the traditional 
Derjaguin-Landau-Verwey-Overbeek (DLVO) theory \cite{dlvo}, the 
effective interactions (solvent and microion mediated) 
between two like-charged colloidal particles are expected to be repulsive
\cite{Chakra}, whereas experiments \cite{Kepler} and numerical
simulations \cite{Gronbech} provide evidence for attraction,
particularly within confined geometries (e.g. when the two macroions are 
close to a charged wall or
between two glass plates).

We consider a mixture of $N$ microions, where species $\alpha$
has charge number $z_\alpha$ and local density $c_\alpha({\bf r})$. 
Disregarding the molecular nature of the solvent, considered to be
a confined electrolyte of dielectric permittivity $\varepsilon$,
the effective interactions between two like-charge colloids
are then analyzed within a generic 
local density functional theory. We show that the 
convexity of the free energy density with respect to the
microions densities is a sufficient condition
for having repulsive effective interactions and
for preventing charge inversion. 
The physical implications of this result are considered in the
final discussion (in particular the repulsion ``often'' follows
from the thermodynamic stability criterion of the corresponding
uncharged mixture of micro-species).
The proof makes use of the results established in 
Refs. \cite{Neu,Sader,PRE}, and considers a geometry which encompasses many
cases of experimental relevance: the confining region
${\cal R}$ is a cylinder of arbitrary cross section 
and length $2L$ (in the limit of large $L$) while
the medium ${\cal R}'$ outside ${\cal R}$ 
is a dielectric continuum of permittivity $\varepsilon'$ 
with the possibility of a uniform
density of surface charges $\sigma$ on the boundary $\partial {\cal R}$.
The standard Neuman 
and Dirichlet boundary conditions 
(constant normal electric field and constant potential respectively)
belong
to the above class, and the dielectric medium in ${\cal R}'$ may contain
an electrolyte solution (this possibility was not considered in
Refs \cite{Neu,Sader}). 
We assume global electroneutrality
in both ${\cal R}$ and ${\cal R}'$
and the two macroions we shall focus on
may be of arbitrary shape provided
the electrostatic potential possesses mirror symmetry with respect
to a plane $Oxy$ between the macroions. 
Our approach holds
irrespective of the specific boundary conditions to be applied on the
macroions (constant charge or constant potential) and is independent
of the sign of the surface charge $\sigma$.

Omitting, for the sake of simplicity, their temperature dependence,
the local \cite{comment} density free energy functionals \cite{Hansen} 
considered here are of the form 
\begin{equation}
{\cal F}(\{c_\alpha\}) = \int_{\cal R} f(\{c_\alpha({\bf r})\})\,d{\bf r} \,+\, 
\frac{1}{2} \, \int_{\cal R} \rho_c({\bf r})\, G({\bf r}, {\bf r}') \, 
\rho_c({\bf r}')\, d{\bf r}\, d{\bf r}',
\label{eq:lda}
\end{equation}
where the Helmholtz free energy density $f$ is a function of all $N$ 
microions densities and the 
local charge density includes the contributions from the
macroions $q_{_{\cal M}}({\bf r})$ as well as from the microions:
\begin{equation}
\rho_c({\bf r}) = \sum_{\alpha=1}^N  z_\alpha\, e\, c_\alpha({\bf r}) \,+\,
q_{_{\cal M}}({\bf r}). 
\label{eq:chargedens}
\end{equation}
In Eq. (\ref{eq:lda}), $G({\bf r},{\bf r'})$ denotes the Green's function such
that the electrostatic potential $\psi({\bf r})$,
which is a solution of Poisson's equation
$\nabla^2 \psi = -(4\pi/\varepsilon) \rho_c({\bf r})$ with the required boundary
conditions on the surface $\partial {\cal R}$, can be cast in the form
\begin{equation}
\psi({\bf r}) = \int_{\cal R} \rho_c{(\bf r'}) G({\bf r},{\bf r'})\, d{\bf r'}.
\end{equation}
$G$ reduces to the usual Coulomb potential in the absence of confinement.
The density functional formulation provided by Eq. (\ref{eq:lda}) encompasses
a large class of modelizations, including {\em a)} 
the standard Poisson Boltzmann theory
when the ideal gas expression 
$f=\sum_\alpha kT\, c_\alpha [\ln(c_\alpha \Lambda_\alpha^3)-1]$
is chosen for the free energy density,
{\em b)} modified Poisson Boltzmann theories derived to account 
for steric effects \cite{Eigen} or more general non-electrostatic interactions
\cite{Lue}, and {\em c)} recent improvements over standard 
mean-field approximations incorporating ion-ion correlations
in the form of a One Component Plasma correction 
%(talk about structuring catastrophe??) 
\cite{Barbosa}. This last case involves a free energy density
which does not only depend on the local densities of microions but also on the
elementary charge $e$, and will be examined in the concluding discussion. 
In general, the present analysis allows to incorporate correlation or non
mean-field contributions, provided these effects translate into a local
free energy density.

Minimization of the functional (\ref{eq:lda}) subject to the normalization 
constraints $\int_{\cal R} c_\alpha({\bf r}) d{\bf r} = N_\alpha$ yields the 
implicit relation between the densities $c_\alpha({\bf r})$ and the electrostatic 
potential:
\begin{equation}
\frac{\partial f}{\partial c_\alpha} \,+\, e \, z_\alpha \psi({\bf r}) \, = 
\mu_\alpha^*, \qquad 1 \leq \alpha \leq N
\label{eq:station}
\end{equation}
where $\mu_\alpha^*$ is the (electro)chemical potential of species $\alpha$.
The effective force acting on a colloid is obtained from integration
of the generalized Maxwell stress tensor $\bbox{\Pi}$ over the surface $S$ of the
macroion:
\begin{equation}
{\bf F} = \oint_S \bbox{\Pi} \cdot {\bf n}\, dS,
\label{eq:forcedef}
\end{equation}
where the unit vector ${\bf n}$ points outward from the surface of integration. 
The local stress tensor follows from the computation of the reversible work
required to locally deform the shape of the confining cylinder, with the result
\begin{equation}
\bbox{\Pi}= -\left[ P + \frac{\varepsilon}{2} 
\,\bbox{\nabla}\psi\cdot\bbox{\nabla}\psi \right]
\bbox{I} + \,\varepsilon\,\bbox{\nabla}\psi \otimes\bbox{\nabla}\psi,
\label{eq:ptensor}
\end{equation}
where the osmotic pressure $P(\{c_\alpha\})$, which depends on position
via the local densities $c_\alpha({\bf r})$, is the pressure that would 
be found in the medium in the absence of an electric field,
{\em with the same densities} $c_\alpha({\bf r})$ \cite{Landau},
%$P$ thus has the same dependence on the $c_\alpha$ as in the uncharged case,
and can be written as the Legendre transform of the free energy density:
\begin{equation}
P(\{c_\alpha\})= -f(\{c_\alpha\}) \,+\, \sum_{\alpha=1}^Nc_\alpha 
\frac{\partial f}{\partial c_\alpha}.
\label{eq:legendre}
\end{equation}
For any stress tensor of the form (\ref{eq:ptensor}), it has been shown in
Refs. \cite{Neu,Sader,PRE} that the effective force ${\bf F}$ given by
Eq. (\ref{eq:forcedef}) has a component $F_z$ along the axis of the cylinder
that can be cast in the form
\begin{eqnarray}
 F_z &=& \int_{Oxy} \left[ P(\psi)_{z=0} - P(\psi)_{z=L} 
- \left(   \psi_{z=0} - \psi_{z=L}\right) 
\frac{\partial P}{\partial \psi}( \psi_{z=L})
\right] dx dy \nonumber\\
&+& \frac{\epsilon}{2} \int_{Oxy} \left( {{\bbox{\nabla} \psi}}_{z=0} - 
{{\bbox{\nabla}\psi}}_{z=L}\right)^2 dx dy,
\label{eq:final}
\end{eqnarray}
where $\epsilon$ denotes $\varepsilon$ (respectively $\varepsilon'$) for the
part of the symmetry plane, denoted $Oxy$, belonging to ${\cal R}$ (respectively
${\cal R}'$). In Equation (\ref{eq:final}), it is understood
that the pressure depends on the potential via 
Eqs. (\ref{eq:legendre}) and (\ref{eq:station}), the latter
giving sense to the notation $\partial P/\partial \psi$.
The subsequent analysis is devoted to the proof that $P$ is a convex-up
function of $\psi$ provided that the free energy density is itself
convex-up with respect to the densities $\{c_\alpha\}$.
Remembering the orientation convention chosen in Refs. \cite{Neu,Sader,PRE},
this in turn establishes the repulsive nature of the effective interactions
($F_z > 0$). From the direct differentiation of Eq. (\ref{eq:legendre})
we get 
\begin{equation}
\frac{\partial P}{\partial \psi} = -\rho_c({\bf r}),
\label{eq:prhoc}
\end{equation}
where use was made of the stationary condition (\ref{eq:station}).
The relation  (\ref{eq:prhoc}) can be recovered from the mechanical equilibrium
condition of a fluid element of microions:
the balance between the electric force and the osmotic constraint
acting on such a fluid element located at point ${\bf r}$ can indeed
be written
\begin{eqnarray}
&&-\bbox{\nabla} P \,=\, \rho_c({\bf r}) \,\bbox{\nabla} \psi 
\label{eq:equil}\\
&\Leftrightarrow & \,\, \left[\frac{\partial P}{\partial \psi} + \rho_c\right]
\bbox{\nabla}\psi = {\bf 0}.
\end{eqnarray}
Equation (\ref{eq:prhoc}) equivalently implies that the tensor $\bbox{\Pi}$ is 
divergence free, as already invoked in Ref. \cite{PRE}
to obtain the osmotic equation of state in the specific case of a 
Modified Poisson Boltzmann theory. 

The second derivative of the osmotic pressure,
\begin{equation}
\frac{\partial^2 P}{\partial \psi^2} \,=\, -e\,\sum_\alpha \, z_\alpha \,
\frac{\partial c_\alpha}{\partial \psi} 
\label{eq:secondder}
\end{equation}
involves the quantities $\partial c_\alpha/\partial \psi$ 
which may be determined from
the condition (\ref{eq:station}):
\begin{equation}
\sum_{\beta=1}^N \,\frac{\partial^2 f}{\partial c_\alpha \partial c_\beta} \,
\frac{\partial c_\beta}{\partial \psi} \,=\, -e \, z_\alpha.
\label{eq:sumbeta}
\end{equation}
Assuming the free energy density to be a convex-up function of the
$\{c_\alpha\}$ implies that the matrix  $\bbox{H}$ with elements
\begin{equation}
H_{\alpha \beta} \equiv \frac{\partial^2 f}{\partial c_\alpha \partial c_\beta}
\end{equation}
is positive definite. Its inverse $\bbox{H}^{-1}$ is consequently 
positive definite from which we can invert relation
(\ref{eq:sumbeta}):
\begin{equation}
\frac{\partial c_\alpha}{\partial \psi} = -e \,\sum_{\beta=1}^N 
\left(\bbox{H}^{-1}\right)_{\alpha\beta} z_\beta.
\label{eq:inverted}
\end{equation}
Upon substitution of Eq. (\ref{eq:inverted}) into (\ref{eq:secondder}) we finally 
obtain
\begin{equation}
\frac{\partial^2 P}{\partial \psi^2} \,=\, e^2\,\sum_{\alpha, \beta} 
\,\left(\bbox{H}^{-1}\right)_{\alpha\beta} z_\alpha\,z_\beta \,\,>\,0.
\label{eq:result}
\end{equation}
The pressure $P$ is therefore a convex-up function of the electrostatic potential,
which establishes the aforementioned connection between 
the convexity of $f$ and the sign of the polyion-polyion effective interactions.
Note that the non convexity of the free energy density 
is a necessary though insufficient condition for
the existence of attraction within a local density approximation formalism.

We now address the question of charge reversal, where the nominal charge
of a bare polyion can be over-screened by its condensed counterions.
This interesting phenomenon (occurring for instance in DNA salt systems
\cite{Shklo}) requires, from Gauss's theorem, the vanishing of the normal
component of the electric field 
($-{\bf n}\cdot \bbox{\nabla} \psi$)
on a closed surface with normal ${\bf n}$
around the polyion. Taking the gradient of 
Eq. (\ref{eq:station}) then implies, without any symmetry assumption
\begin{equation}
\sum_{\beta=1}^{N} \,\frac{\partial^2 f}{\partial c_\alpha \partial c_\beta} \,
\left[\bf{n}\cdot\bbox{\nabla} c_\beta\right] = 0,
\end{equation}
hence the existence of a vanishing eigenvalue [with eigenvector
$({\bf n}\cdot \bbox{\nabla} c_\beta)_{\beta=1\ldots N}$]
of the stability matrix $\bbox{H}$,
which is impossible under the assumption of positive definiteness.

Within the framework of Poisson Boltzmann theory,
the repulsive nature of the effective electrostatic force between two similar
colloidal bodies \cite{Neu,Sader,PRE}
and the impossibility of charge reversal
immediately follows from the previous analysis.
However, if for the sake of analytical tractability and
in the spirit of Ref. \cite{Goulding},
the ideal free energy density is further Taylor expanded
up to order $\cal N$ around the mean densities $\{\overline{c}_\alpha\}$, 
we obtain
\begin{equation}
H_{\alpha \beta} = \frac{\delta_{\alpha\beta}}{\overline{c}_\alpha}\,
\sum_{i=0}^{{\cal N}-2} \left( 1-\frac{c_\alpha}{\overline{c}_\alpha}
\right)^i = 
\left[1-\left(1-\frac{c_\alpha}{\overline{c}_\alpha}\right)^{{\cal N}-1}\, \right]
\frac{\delta_{\alpha\beta}}{c_\alpha}.
\end{equation}
The truncated Hessian $\bbox{H}$ is therefore positive definite 
and gives a repulsive force for even 
values of ${\cal N}$ (as in the non-linear Poisson-Boltzmann theory),
while it can be non-positive definite and the force attractive 
for odd orders ${\cal N}$ 
provided the local density of one counterion species exceeds twice its mean density.
This explains the repulsion within linearized Poisson-Boltzmann 
theory \cite{Goulding2} 
(where linearization amounts to taking ${\cal N}=2$ \cite{Hansen}), 
and the attraction seen in Ref. \cite{Goulding} with an 
expansion pushed one order further and truncated
after third order to account for the triplet interactions between two polyions
and a charged wall.

We have shown that effective attractive pair potentials and over-charging 
are both 
ruled out under the assumption of positive definiteness for the stability matrix
$\bbox{H}$. This is a necessary requirement when the free energy density 
does not depend on the elementary charge $e$. 
The key ingredient here is the existence of 
a {\em thermodynamically stable} neutral mixture of micro-species 
described by the same density $f\{c_\alpha\}$ as its charged counterpart. 
On the other hand, the inclusion of correlations and/or 
fluctuations corrections is believed to be 
essential in order to describe the appearance of attractive interactions. If their
correction to the mean-field electrostatic term 
$\int \rho_c G \rho_c$ in the right hand
side of Eq. (\ref{eq:lda}) is taken 
into account, and translated into an $e$-dependent 
free energy density \cite{Barbosa,Diehl}, 
the thermodynamic stability condition is 
now the positive definiteness of the integral operator
whose kernel is defined by 
\begin{equation}
\frac{\partial^2 f}{\partial c_\alpha \partial c_\beta}\, \delta({\bf r}-{\bf r}')\,
+ e^2 \, z_\alpha \,z_\beta\, G({\bf r},{\bf r}'),
\end{equation}
and it is in general impossible to find a {\em locally}\/ neutral mixture 
described by the same density $f$. $\bbox{H}$ need not therefore 
be positive definite. However, non convex-up densities $f$ such 
as the correction put forward by Stevens and Robbins \cite{Stevens} 
by extrapolation of the
Monte Carlo data for the homogeneous One Component Plasma
lead to an instability at high densities (the so called structuring catastrophe), 
and cannot be used in a thermodynamically stable way within a local density
functional theory \cite{Barbosa,Diehl}. 
The recent attempt of Barbosa {\it et al.} \cite{Barbosa}
to circumvent
this instability gives rise to a convex-up $f$, with results 
reproducing satisfactorily the ionic correlations present in Molecular
Dynamics simulations.
The present work 
indicates that the effective interactions are then necessarily repulsive
and charge inversion impossible, regardless of the analytical 
complexity of the formalism. 

To conclude, repulsive effective pair potentials seem generic
within local density approximations.
The description of attraction and over-charging thus
requires involved density functional theories, and it is not sufficient
to incorporate correlation effects in a local formulation.
Our analysis points
to the importance of non-local effects, that can be accounted for 
by weighted density functionals \cite{Stevens,Diehl}, or simpler approaches
relying on non-electrostatic depletion interactions \cite{Tamashiro}.

%\bigskip
The author would like to thank J.L. Raimbault, D. Levesque,
J.P. Hansen, B. Jancovici, H. Hilhorst, M. Deserno 
and D. Rowan for helpful discussions.

%\end{multicols}


\begin{thebibliography}{99}
\bibitem{comment} 
        The ``locality'' does not refer to the van der Waals like mean-field
	term $\int \rho_c G \rho_c$ but to the term $\int\! f$ that may contain
	correlation contributions besides the usual ideal part \cite{Hansen}.
\end{thebibliography}
\end{document}